# Enhancing Decision Support in Construction through Industrial AI

Parul Khanna
Department of Civil, Environmental and Natural Resources Engineering
Luleå University of Technology
0920-492386

parul.khanna@ltu.se

Sameer Prabhu
Department of Civil, Environmental and Natural Resources Engineering
Luleå University of Technology
0920-492198

sameer.prabhu@ltu.se

Ramin Karim
Department of Civil, Environmental and Natural Resources Engineering
Luleå University of Technology
0920-492344

ramin.karim@ltu.se

Phillip Tretten
Department of Social Sciences, Technology and Arts
Luleå University of Technology
0920-492855

phillip.tretten@ltu.se

## ABSTRACT

The construction industry is presently going through a transformation led by adopting digital technologies that leverage Artificial Intelligence (AI). These industrial AI solutions assist in various phases of the construction process, including planning, design, production and management. In particular, the production phase offers unique potential for the integration of such AI-based solutions. These AI-based solutions assist site managers, project engineers, coordinators and other key roles in making final decisions. To facilitate the decision-making process in the production phase of construction through a human-centric AI-based solution, it is important to understand the needs and challenges faced by the end users who interact with these AI-based solutions to enhance the effectiveness and usability of these systems. Without this understanding, the potential usage of these AI-based solutions may be limited. Hence, the purpose of this research study is to explore, identify and describe the key factors crucial for developing AI solutions in the construction industry. This study further identifies the correlation between these key factors. This was done by developing a demonstrator and collecting quantifiable feedback through a questionnaire targeting the end users such as site managers and construction professionals. This research study will offer insights into developing and improving these industrial AI solutions, focusing on Human-System Interaction aspects to enhance decision support, usability, and overall, AI solution adoption.

## 1. INTRODUCTION

The construction industry adheres to a well-defined and systematic phase to ensure project success. For informed decision making and resource allocation, project leader depends on data. Although new advanced technologies are being integrated in the construction process, their adoption remains limited. However, more companies are now opting for advanced technologies including AI and automation to have a competitive advantage in the industry [1]. The success of a construction project is usually measured by four indicators including cost, schedule, quality and safety [2]. The construction industry follows a structured process which can be divided into four phases [3], shown in **Error! Reference source not found.**. Each phase plays a crucial role in successful execution of the projects. The first phase is the planning or the initiation phase, during which project objectives, feasibility assessment, budget proposals and scheduling are performed. It is essential for establishing project directions and to prevent delays and budget overruns. During the design phase, blueprints and specifications are created in accordance with the building codes and safety regulations. The production phase includes the execution of actual building activities, involving tasks from material logistics to safety compliance. This phase is labour-intensive and requires real-time decision-making. The long-term sustainability and operational efficiency of the structure is ensured through the management phase.

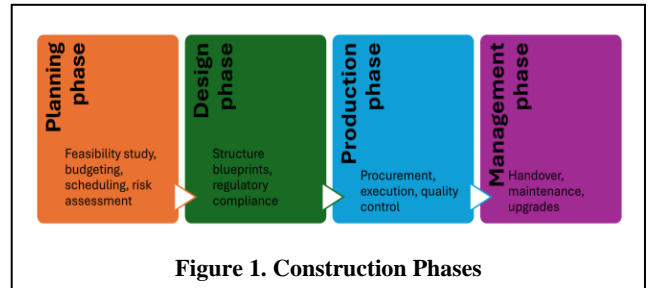

**Figure 1. Construction Phases**

For the scope of this study, we focus on the production phase as it emerges as a critical phase where decision-making significantly impacts the project outcomes. This phase is often complex in nature as it includes material handling, resource management, equipment operations, and safety compliance. Accurate and timely decisions are essential in this phase to control the cost and ensure project deadlines and the safety of on-site activities. Therefore, enhancing decision support during this phase is essential to improve the overall project performance.

Artificial Intelligence (AI) term was introduced in the 1950's and since then it had seen shifts in the level of interest by scholars and practitioners [4], [5]. Advanced digital technologies, particularly AI are transforming industries, and have successfully been utilized to enhance efficiency, safety and security. However, along with benefits of AI applications, certain challenges relevant exist in the construction industry [6]. AI is developing as a revolutionary force across all the phases of construction. The production phase presents unique opportunities and challenges for the integration of AI-based solutions. On a construction site, the site manager faces various dynamic and unpredictable situations requiring informed decision support. AI-based solution has the potential to assist the decision-making by analysing and identifying patterns within the data. Despite the potential, the adoption of AI solutions remains limited due to challenges like trust, security, expert shortages, and computing power requirements [6]. A successful adoption and integration of AI requires a fundamental understanding of both functional and non-functional requirements. This requirement analysis ensures the identification of both technological capabilities and human-system interaction (HSI) needs of the system.

This paper explores the functional and non-functional requirements of the users of AI-solution by understanding their insights and feedback. This is to ensure their needs and challenges are addressed during the development of the AI system and to enhance the Human-System Interaction. The technical capability of AI is emphasised when developing the solution, often overlooking the human-centric design principles necessary for a successful and effective adoption. To bridge this gap, this research aims to identify and describe the key factors and their correlations when developing industrial AI solutions that are both technically robust and user-friendly. Focusing on end-user perspectives and collecting quantifiable feedback through a demonstrator and questionnaire, this study can provide a foundation for creating an industrial AI system.

## 2. Background

Digitalization and AI technologies in the construction industry have been adopted to enhance productivity, optimize operations, improve site safety and security [7]. However, [6] highlights that the construction industry lags behind compared to the manufacturing and telecommunication sectors. AI in the construction industry is used to automate planning and scheduling, safety management, smart construction [6]. [8] focused on technical capabilities of AI, including machine learning and big data for large datasets predicting delays, optimize scheduling. Machine learning to help in safety and risk management, AI-powered generative design. A comprehensive review of applications of AI in the construction industry is given in [7]. Machine learning algorithms like neural networks, and support vector machines, were used for cost predictions, and risk analysis whereas for safety assessment and decision support expert systems, rule-based systems were mentioned [7]. The impact of AI on project management, risk management, cost control and scheduling were discussed in [9]. AI enables optimized resource allocation, real-time monitoring and predictive analytics [9]. For automated code compliance checks, Natural Language Processing is being tested to extract information for regulatory texts [10]. From the past construction management project documents AI could enhance information retrieval hence facilitating decision-making easier and faster [4], [11]. AI technologies like machine learning and artificial neural networks possess the capability to manage explicit and tacit knowledge in construction projects for the encoding of visual building information [4], [10].

Despite the benefits, adopting AI in construction is challenging. Due to high complexity and fragmented workflows, the adoption of AI is limited [9]. The high initial costs, limited digital infrastructure, and lack of skilled AI professionals in the construction sector are some other challenges mentioned in [9]. Construction managers find AI solutions too complex and difficult to understand, leading to low adoption [12]. As the existing workflow relies on manual processes and legacy systems, there is resistance to change from the traditional system [13]. Construction projects involve multiple stakeholders, making data integration difficult; uncertain guidelines on AI applications create uncertainty; data security and privacy create additional barriers [14].

[15] acknowledges the importance of requirement analysis when implementing performance-based building (PBB) approaches to enhance innovation in the construction industry. [16] suggested that successful AI adoption requires emphasis on Human-System Interaction, indicating that AI solutions should be designed with the end-user's workflow in mind rather than only technical innovation. To facilitate AI adoption [16], [10] explore approaches emphasizing usability, decision support and human-AI collaboration. Challenges like user acceptance, trust, and interface complexity have frequently left construction professionals often reluctant to integrate AI-based tools [17]. AI solution is more likely to have higher usability and acceptance if it integrates seamlessly into existing workflows [17]. In order to facilitate AI adoption, several key indicators have been suggested, to increase trust in AI systems, transparency, and explainability should be increased [18], [10]. The construction site has dynamic conditions, and AI tools should be adaptable and evolve based on user experience. Human experience remains crucial for complex decision-making, whereas AI systems are best suited for repetitive tasks. [18] recommends a hybrid AI system that integrates AI-driven insights with human judgement. Organizations must prioritize the reskilling and upskilling of personnel to facilitate collaboration with AI systems. To enhance productivity, and improve decision-making, human-AI collaboration needs to be implemented [18].

## 3. Research Methodology

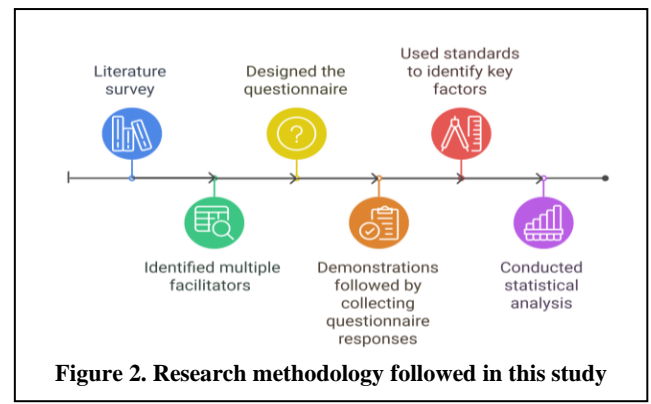

**Figure 2. Research methodology followed in this study**

This study combined qualitative and quantitative research approaches to ensure a comprehensive exploration. We opted for a mixed-method approach as it was suitable for analysing both, quantitative measures and subjective experiences. This provided us with a holistic understanding of the analysis. Our research methodology included a literature survey, demonstrations followed by a questionnaire, semi-structured interviews and thematic analysis to identify the key factors to facilitate the adoption of AI and hence enhance decision support in construction.

### 3.1 Literature Survey

The study began with an exploratory literature review to provide a theoretical foundation. This was done to identify relevant research trends, gaps, and key factors related to the adoption and success of technological tools including AI to enhance decision support in the construction industry. The literature search focused on the abstracts of articles within the 2022-2025 period, and the searched keywords included "construction industry" AND "decision support". The search provided 83 articles from Scopus and 66 articles from Google Scholar. Duplicate articles from both searches were removed, and a scope assessment was done to avoid unrelated articles. Additionally, a backward citation analysis was done. Articles from the initial search were reviewed for their reference lists to identify additional relevant articles. A total of XX relevant articles were reviewed in this study.

### 3.2 Demonstrations and Questionnaire

Based on the gaps and factors identified from the literature survey, a questionnaire was developed in reference to a demonstrator, showcasing the use of Industrial AI in the construction industry. The developed questionnaire was structured into two types of questions and supported English and Swedish languages. Following the work of [19], which states that Likert-scale questions are commonly used to measure perceptions, and experiences of participants in a survey due to their ability to capture varying degrees of opinion and ease of analysis. Therefore, most of the designed questions were on a Likert scale with a few on 5-point scale (ranging from "Strongly Disagree" to "Strongly Agree") and a few on 10-point scale (ranging from rating 1: Not at all likely to 10: Extremely likely). However, a few questions included text fields to collect user suggestions and feedback. To ensure a clear understanding among the participants, a presentation on a developed demonstrator was given followed by a hands-on experience with the AR/VR glasses to interact with the developed tool. The developed demonstrator illustrated the potential use of Industrial AI for assisting site managers as a decision support tool. It was developed in reference to a scaffolding on a construction site. These demonstrations explained the system's workflow, emphasizing the key stages starting with data acquisition, data processing, and finally, visualization. The demonstration consisted of the following steps:

Data Acquisition:

This step explained the use of a handheld LiDAR scanner to scan and collect the 3D point cloud data from a construction site, a scaffolding in this case. LiDAR technology was selected in developing this demonstrator due to its precision in capturing 3D point cloud data.

Data Processing:

This step involved processing the raw data from the LiDAR scanner and optimizing it to reduce computational load using data filtering and point cloud compression. It enhanced the computational efficiency of the scanned model enabling their smooth rendering in the AR/VR glasses.

Visualization in AR/VR:

In this step, the processed point cloud model was rendered using Azure Remote Rendering services in AR/VR glasses. The digital model simulated the real-world construction site. This was done to demonstrate the practicality of the technology and to highlight the potential use of such tools for remote collaboration. It also highlighted how intuitive and interactive systems facilitate a deeper understanding of its capabilities in construction workflows.

The demonstrator was developed not only to showcase the potential applications of the technology but also to encourage the participants to brainstorm and reflect on the challenges associated with its implementation in real-world construction scenarios.

Nineteen participants filled out the questionnaire. The participants were intentionally selected from a diverse range of backgrounds and experience levels with respect to the construction industry and experience with technologies. The participants consisted of 15 males and 4 females, with a wide range of ages and professional backgrounds. Nine participants were between 18-25 years, two were in the 26-35, 36-45, and 46-55 age groups, one was over 55 years old, and one participant preferred not to disclose their age. The group comprised 10 final-year construction students with on-site experience and 9 professionals with a mix of academic and industrial backgrounds with respect to the construction industry. This group of professionals included researchers, professors, site managers, production managers, and branch and association managers specializing in construction and technology integration.

### 3.3 Semi-structured interviews

Semi-structured interviews were chosen to gather qualitative insights from the participants. This approach was chosen because it provides flexibility in exploring predetermined topics while allowing the possibility of open-ended discussions with the participants [20]. The interview questions focused on the participant's insights on the production phase of the construction process. The discussion included the intention of understanding the role of humans in the subprocess in the production phase and their insights on having an AI-based system to assist humans in decision-making. The interviews were also directed towards gaining their suggestions in areas where AI could help and if so, what could be the limitations in its successful adoption. Additionally, participants were also asked to discuss the areas where they think traditional practices were enough and adopting AI solutions would not be very beneficial. The interviews also focused on gaining insights from the participants on the limited adoption of AI in general in the construction industry.

### 3.4 Data Analysis: Thematic Analysis

The data collected from the literature review, questionnaire responses and insights from the semi-structured interviews were analysed using a thematic analysis approach. Thematic analysis was chosen for this study as it is a commonly used qualitative method for identifying, analysing, and identifying themes or patterns in data [21]. After careful analysis and interpretation of the collected data, findings were aligned with the relevant standards [22] to identify the key factors.

# 4. RESULTS
This section reveals the key insights into the perceptions of incorporating AI in enhancing decision-making in the production

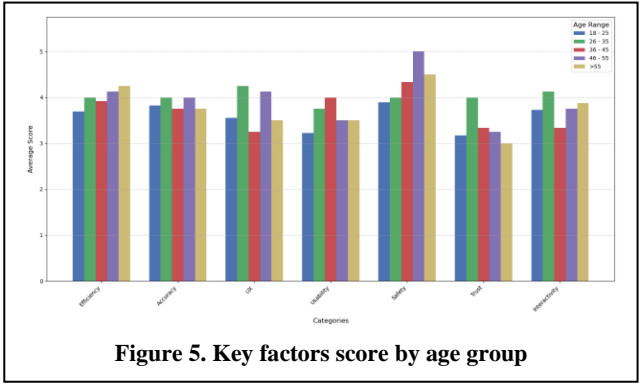

**Figure 5. Key factors score by age group**

phase. The results are broadly classified into 4 headings as discussed below:

## 4.1 Identifying key factors
Using established standards and thematic analysis, discussed in section 3.4, key factors which are required to incorporate AI to enhance decision-making are identified. These indicators include efficiency, accuracy, user experience (UX), trust, usability, safety, and interactivity. These key factors impact both system performance and user acceptance and are critical for the successful adoption of AI. These factors comprise the functional and non-functional requirements. Accuracy, safety, and interactivity could be identified as functional requirements. These factors define the essential capabilities that the system must have. Meanwhile, efficiency, user experience (UX), usability, and trust could be classified as non-functional requirements. These factors focus on how the system operates and supports user interaction. Optimizing these key factors while developing AI solutions can enhance decision-making.

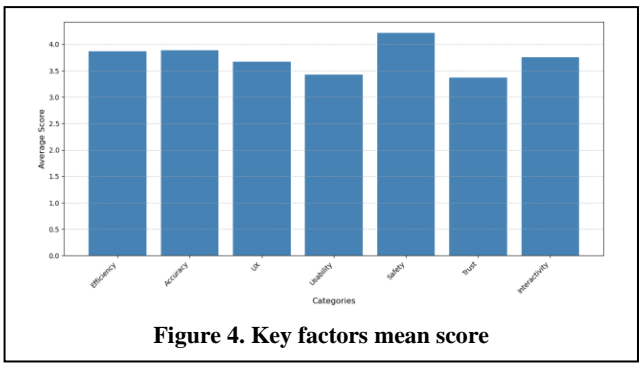

**Figure 4. Key factors mean score**

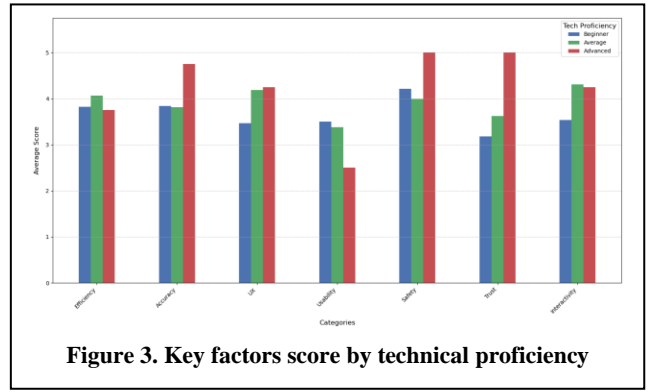

**Figure 3. Key factors score by technical proficiency**

## 4.2 Score across key factors
The average score for various key indicators is analysed and shown in Figure 3, emphasizing the critical role, safety received the highest score. Trust scored lower than other key factors, reflecting concerns about the confidence and reliability of the AI solution. Other key factors have a moderate score, suggesting future potential with further improvements and optimizations.

## 4.3 Variations in key factor scores
In this section, the scores for key factors are further analysed based on professional and demographic differences.

### 4.3.1 Scores by age group
As shown in Figure 4, participant's perception varies across different age groups.

Safety and efficiency are rated high by participants over 55 years of age; however, they gave lower scores in trust, usability and interactivity, suggesting hesitancy toward adopting new technologies. Given the sample size, the participants provided moderate scores across all categories, indicating a balance between openness to technology and caution. However, during the semi-structured interviews, safety was emphasized as the primary concern, which is also reflected through data analysis. Middle-aged participants are more receptive and comfortable with technology compared to the older age group.

### 4.3.2 Scores by technical proficiency
Figure 5 illustrates the relationship between technical proficiency and key factor scores. Participants with advanced proficiency indicate confidence in technology's capabilities reflected through high scores in accuracy and interactivity. Beginner-level proficiency gave lower scores on trust and usability, suggesting a limited understanding of AI capabilities.

### 4.3.3 Scores by profession
The key factor scores between students and professionals are shown in Figure 6.

Professionals rated safety a high score, reflecting their focus on risk management and reliable operations whereas students showed greater interest in the ease of use and engagement provided by technology gave higher scores for the interactivity key factor.

### 4.4 Issues and suggestions for site managers
A word cloud is presented in Figure 7, summarizing the issues and suggestions mentioned by the participants. Maintaining safety working conditions is a critical concern, effective planning, regular inspection rounds and managing dependencies are seen as elements to reduce delays and conflict on site.

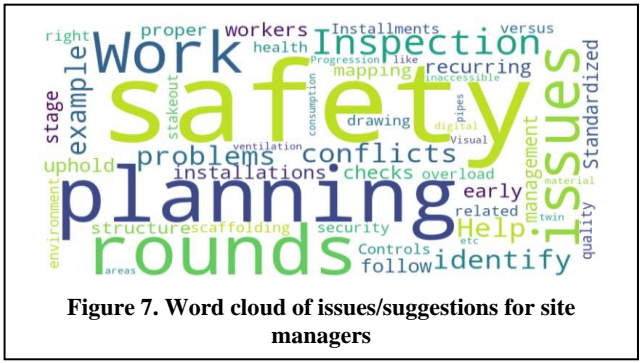

Figure 7. Word cloud of issues/suggestions for site managers

## 5. CONCLUSION
The key factors influencing the adoption and acceptance of technology including AI in construction's production phase are identified in this paper. These factors were categorized into functional and non-functional requirements. Accuracy, safety, and interactivity were identified as functional requirements, defining the essential tasks and core operations the system must perform. In contrast, efficiency, UX, usability, and trust were classified as non-functional requirements, emphasizing how the system should operate to support user engagement and effective interaction. As reflected by both data analysis and semi-structured interviews, safety is the top priority. AI solutions that prioritize safety when dealing with critical and high-stakes tasks are likely to gain trust and acceptance. Particularly in the older generation, the key factor of trust remains a significant barrier to AI adoption. However, increased transparency, gradual integration of automation, tailored training and support programs can help build confidence in the users. The generational and proficiency-based difference in perceptions towards AI is also revealed in this study. Greater acceptance of technology, focusing on interactivity is exhibited in younger and technically advanced participants, whereas older and less proficient participants prioritized safety and expressed hesitance in AI adoption. These findings contribute to enhancing decision-making processes and facilitating human-system interaction (HSI). This can ensure that the digital tools are user-friendly and intuitive to use. Improved HSI enhances usability and interactivity, helping users feel more engaged and confident in using AI-based systems. This interaction fosters trust and adoption, especially in environments involving complex tasks that require smooth collaboration between human and automated systems. Additionally, these findings can serve as a valuable base for a framework for assessing the technology readiness level (TRL), of AI and digital systems in construction workflows to evaluate the maturity of a technology.

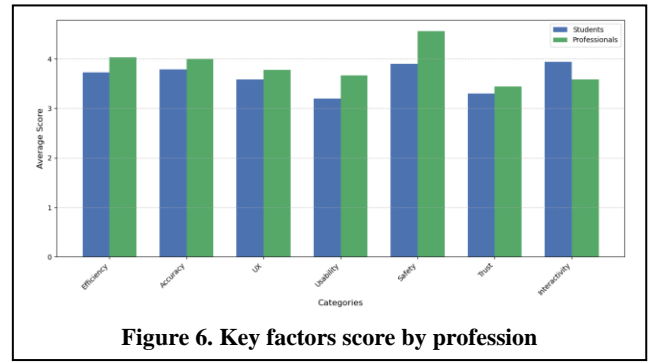

Figure 6. Key factors score by profession

## 6. ACKNOWLEDGMENTS
The authors gratefully acknowledge the European Commission for supporting the Marie Sklodowska Curie program through the H2020 ETN MOIRA project (GA 955681). The authors would also like to thank NCC, HÖ Allbygg, Byggföretagen, SBUF, Smart Built Environment and Formas for their support and contributions. We extend our appreciation to Jarkko Erikshammar, senior lecturer at LTU, for his outreach support. Lastly, we also thank all questionnaire participants for their valuable contributions to this study.